\documentclass[aps,prl,twocolumn,superscriptaddress,showpacs]{revtex4-1}

\usepackage{amsmath}
\usepackage{graphicx}
\usepackage{pifont}
\usepackage{array}
\usepackage{rotate}
\usepackage{vmargin}
\usepackage{lineno}
\usepackage{fancyhdr}
\usepackage{color}
\usepackage{url}
\usepackage{wasysym}
\usepackage{textcomp}

\begin{document}


\sloppy


\newpage

\title{Search for magnetic monopoles with the MoEDAL forward trapping detector in 13 TeV proton-proton collisions at the LHC}

\author{B.~Acharya}
\affiliation{Theoretical Particle Physics \& Cosmology Group, Physics Dept., King's College London, UK}
\affiliation{International Centre for Theoretical Physics, Trieste, Italy}   

\author{J.~Alexandre}
\affiliation{Theoretical Particle Physics \& Cosmology Group, Physics Dept., King's College London, UK}

\author{S.~Baines}
\affiliation{Formerly at School of Physics and Astronomy, The University of Manchester, UK -- Associate member}

\author{P.~Benes}
\affiliation{IEAP, Czech Technical University in Prague, Czech~Republic}

\author{B.~Bergmann}
\affiliation{IEAP, Czech Technical University in Prague, Czech~Republic}

\author{J.~Bernab\'{e}u}
\affiliation{IFIC, Universitat de Val\`{e}ncia - CSIC, Valencia, Spain}

\author{H.~Branzas}
\affiliation{Institute of Space Science, Bucharest - M\u{a}gurele, Romania}

\author{M.~Campbell}
\affiliation{Experimental Physics Department, CERN, Geneva, Switzerland}

\author{L.~Caramete}
\affiliation{Institute of Space Science, Bucharest - M\u{a}gurele, Romania}

\author{S.~Cecchini}
\affiliation{INFN, Section of Bologna, Bologna, Italy}



\author{M.~de~Montigny}
\affiliation{Physics Department, University of Alberta, Edmonton, Alberta, Canada}


\author{A.~De~Roeck}
\affiliation{Experimental Physics Department, CERN, Geneva, Switzerland}

\author{J.~R.~Ellis}
\affiliation{Theoretical Particle Physics \& Cosmology Group, Physics Dept., King's College London, UK}
\affiliation{Theoretical Physics Department, CERN, Geneva, Switzerland}


\author{M.~Fairbairn}
\affiliation{Theoretical Particle Physics \& Cosmology Group, Physics Dept., King's College London, UK}

\author{D.~Felea}
\affiliation{Institute of Space Science, Bucharest - M\u{a}gurele, Romania}

\author{J.~Flores}
\affiliation{Formerly at Department of Physics and Astronomy, Stony Brook University, NY, USA -- Associate member}

\author{M.~Frank}
\affiliation{Department of Physics, Concordia University, Montr\'{e}al, Qu\'{e}bec,  Canada}

\author{D.~Frekers}
\affiliation{Physics Department, University of Muenster, Muenster, Germany}

\author{C.~Garcia}
\affiliation{IFIC, Universitat de Val\`{e}ncia - CSIC, Valencia, Spain}




\author{A.~M.~Hirt}
\affiliation{Department of Earth Sciences, Swiss Federal Institute of Technology, Zurich, Switzerland -- Associate member}



\author{J.~Janecek}
\affiliation{IEAP, Czech Technical University in Prague, Czech~Republic}

\author{M.~Kalliokoski}
\affiliation{Beams Department, CERN, Geneva, Switzerland}

\author{A.~Katre}
\affiliation{Section de Physique, Universit\'{e} de Gen\`{e}ve, Geneva, Switzerland}

\author{D.-W.~Kim}
\affiliation{Physics Department, Gangneung-Wonju National University, Gangneung, Republic of Korea}


\author{K.~Kinoshita}
\affiliation{Physics Department, University of Cincinnati, Cincinnati, Ohio, USA}

\author{A.~Korzenev}
\affiliation{Section de Physique, Universit\'{e} de Gen\`{e}ve, Geneva, Switzerland}

\author{D.~H.~Lacarr\`ere}
\affiliation{Experimental Physics Department, CERN, Geneva, Switzerland}

\author{S.~C.~Lee}
\affiliation{Physics Department, Gangneung-Wonju National University, Gangneung, Republic of Korea}

\author{C.~Leroy}
\affiliation{D\'{e}partement de physique, Universit\'{e} de Montr\'{e}al, Qu\'{e}bec,  Canada}

\author{A.~Lionti}
\affiliation{Section de Physique, Universit\'{e} de Gen\`{e}ve, Geneva, Switzerland}

\author{J.~Mamuzic}
\affiliation{IFIC, Universitat de Val\`{e}ncia - CSIC, Valencia, Spain}

\author{A.~Margiotta}
\affiliation{INFN, Section of Bologna \& Department of Physics \& Astronomy, University of Bologna, Italy}

\author{N.~Mauri}
\affiliation{INFN, Section of Bologna, Bologna, Italy}

\author{N.~E.~Mavromatos}
\affiliation{Theoretical Particle Physics \& Cosmology Group, Physics Dept., King's College London, UK}

\author{P.~Mermod}
\email[Corresponding author:\\]{philippe.mermod@cern.ch}
\affiliation{Section de Physique, Universit\'{e} de Gen\`{e}ve, Geneva, Switzerland}

\author{V.~A.~Mitsou}
\affiliation{IFIC, Universitat de Val\`{e}ncia - CSIC, Valencia, Spain}

\author{R.~Orava}
\affiliation{Physics Department, University of Helsinki, Helsinki, Finland}

\author{B.~Parker}
\affiliation{The Institute for Research in Schools, Canterbury, UK}

\author{L.~Pasqualini}
\affiliation{INFN, Section of Bologna \& Department of Physics \& Astronomy, University of Bologna, Italy}

\author{L.~Patrizii}
\affiliation{INFN, Section of Bologna, Bologna, Italy}

\author{G.~E.~P\u{a}v\u{a}la\c{s}}
\affiliation{Institute of Space Science, Bucharest - M\u{a}gurele, Romania}

\author{J.~L.~Pinfold}
\affiliation{Physics Department, University of Alberta, Edmonton, Alberta, Canada}


\author{V.~Popa}
\affiliation{Institute of Space Science, Bucharest - M\u{a}gurele, Romania}

\author{M.~Pozzato}
\affiliation{INFN, Section of Bologna, Bologna, Italy}

\author{S.~Pospisil}
\affiliation{IEAP, Czech Technical University in Prague, Czech~Republic}

\author{A.~Rajantie}
\affiliation{Department of Physics, Imperial College London, UK}

\author{R.~Ruiz~de~Austri}
\affiliation{IFIC, Universitat de Val\`{e}ncia - CSIC, Valencia, Spain}

\author{Z.~Sahnoun}
\affiliation{INFN, Section of Bologna, Bologna, Italy}
\affiliation{Centre for Astronomy, Astrophysics and Geophysics, Algiers, Algeria}

\author{M.~Sakellariadou}
\affiliation{Theoretical Particle Physics \& Cosmology Group, Physics Dept., King's College London, UK}

\author{S.~Sarkar}
\affiliation{Theoretical Particle Physics \& Cosmology Group, Physics Dept., King's College London, UK}

\author{G.~Semenoff}
\affiliation{Department of Physics, University of British Columbia, Vancouver, British Columbia, Canada}

\author{A.~Shaa}
\affiliation{Formerly at Department of Physics and Applied Physics, Nanyang Technological University, Singapore  -- Associate member}

\author{G.~Sirri}
\affiliation{INFN, Section of Bologna, Bologna, Italy}

\author{K.~Sliwa}
\affiliation{Department of Physics and Astronomy, Tufts University, Medford, Massachusetts, USA}

\author{R.~Soluk}
\affiliation{Physics Department, University of Alberta, Edmonton, Alberta, Canada}

\author{M.~Spurio}
\affiliation{INFN, Section of Bologna \& Department of Physics \& Astronomy, University of Bologna, Italy}

\author{Y.~N.~Srivastava}
\affiliation{Physics Department, Northeastern University, Boston, Massachusetts, USA}


\author{M.~Suk}
\affiliation{IEAP, Czech Technical University in Prague, Czech~Republic}

\author{J.~Swain}
\affiliation{Physics Department, Northeastern University, Boston, Massachusetts, USA}

\author{M.~Tenti}
\affiliation{INFN, CNAF, Bologna, Italy}

\author{V.~Togo}
\affiliation{INFN, Section of Bologna, Bologna, Italy}


\author{J.~A.~Tuszy\'{n}ski}
\affiliation{Physics Department, University of Alberta, Edmonton, Alberta, Canada}

\author{V.~Vento}
\affiliation{IFIC, Universitat de Val\`{e}ncia - CSIC, Valencia, Spain}

\author{O.~Vives}
\affiliation{IFIC, Universitat de Val\`{e}ncia - CSIC, Valencia, Spain}

\author{Z.~Vykydal}
\affiliation{IEAP, Czech Technical University in Prague, Czech~Republic}

\author{T.~Whyntie}
\affiliation{The Institute for Research in Schools, Canterbury, UK}
\affiliation{Queen Mary University of London, London, UK}

\author{A.~Widom}
\affiliation{Physics Department, Northeastern University, Boston, Massachusetts, USA}

\author{G.~Willems}
\affiliation{Physics Department, University of Muenster, Muenster, Germany}

\author{J.~H.~Yoon}
\affiliation{Physics Department, Konkuk University, Seoul, Korea}

\author{I.~S.~Zgura}
\affiliation{Institute of Space Science, Bucharest - M\u{a}gurele, Romania}

\collaboration{THE MoEDAL COLLABORATION}
\noaffiliation

\date{\today}

\begin{abstract}
MoEDAL is designed to identify new physics in the form of long-lived highly-ionising particles produced in high-energy LHC collisions. Its arrays of plastic nuclear-track detectors and aluminium trapping volumes provide two independent passive detection techniques. We present here the results of a first search for magnetic monopole production in 13~TeV proton-proton collisions using the trapping technique, extending a previous publication with 8~TeV data during LHC run-1. A total of 222~kg of MoEDAL trapping detector samples was exposed in the forward region and analysed by searching for induced persistent currents after passage through a superconducting magnetometer. Magnetic charges exceeding half the Dirac charge are excluded in all samples and limits are placed for the first time on the production of magnetic monopoles in 13~TeV $pp$ collisions. The search probes mass ranges previously inaccessible to collider experiments for up to five times the Dirac charge.
\end{abstract}

\pacs{14.80.Hv, 13.85.Rm, 29.20.db, 29.40.Cs}

\maketitle

The existence of a magnetically charged particle would add symmetry to Maxwell's equations and explain why electric charge is quantised in nature, as shown by Dirac in 1931~\cite{Dirac1931}. In addition to providing a consistent quantum theory of magnetic charge and elucidating electric charge quantisation, Dirac predicts the fundamental magnetic charge number (or Dirac charge) to be $g_{\rm D}=\frac{1}{2\alpha_{em}}\simeq 68.5$ where $\alpha_{em}$ is the fine-structure constant. Consequently, in SI units, magnetic charge can be written in terms of the dimensionless quantity $g_{\rm D}$ as $q_m = ng_{\rm D}ec$ where $n$ is an integer number, $e$ is the proton charge, and $c$ is the speed of light in vacuum. Because $g_{\rm D}$ is large, a fast monopole is expected to induce ionisation in matter thousands of times higher than a particle carrying the elementary electric charge. Additionally, the existence of the monopole as a topological soliton is a prediction of theories of the unification of forces~\cite{tHooft1974,Polyakov1974,Scott1980,Preskill1984} where the monopole mass is determined by the mass scale of the symmetry breaking that allows nontrivial topology. For a unification scale of $10^{16}$~GeV such monopoles would have a mass in the range $10^{17}-10^{18}$~GeV. In unification theories involving a number of symmetry-breaking scales~\cite{Lazarides1980,Kirkman1981,Kephart2001} monopoles of much lower mass can arise, although still beyond the reach of the LHC. However, an electroweak monopole has been proposed~\cite{Cho1997,Cho2015,Ellis2016,Cho2016} that is a hybrid of the Dirac and 't Hooft-Polyakov monopoles~\cite{tHooft1974,Polyakov1974} with a mass that is potentially accessible at the LHC.

Monopole relics from the early Universe have been extensively searched for in cosmic rays and in materials~\cite{Burdin2015,Patrizii2015}. In the laboratory, monopole-antimonopole pairs are expected to be produced in particle collisions, provided the collision energy exceeds twice the monopole mass $M$. Each time an accelerator accessed a new energy scale, dedicated searches were made in new monopole mass regions~\cite{Fairbairn2007}. The Large Hadron Collider (LHC) is no exception to this strategy as a comprehensive monopole search programme using various techniques has been devised to probe TeV-scale monopole masses for the first time~\cite{DeRoeck2012a,MoEDAL2014}. The results obtained by MoEDAL using 8 TeV $pp$ collisions allowed the existing LHC constraints on monopole pair production ~\cite{ATLAS2015a} to be improved to provide limits on monopoles with $|g|\le 3g_{\rm D}$ and $M\le 3500$~GeV~\cite{MoEDAL2016}.

In 2015, an increase in the LHC $pp$ collision energy from 8 TeV to 13 TeV was achieved, opening a significant discovery opportunity window. This paper presents the first monopole search results in this new energy regime, using the forward monopole trapping detector of the MoEDAL experiment exposed to $0.371\pm 0.004$~fb$^{-1}$ of 13 TeV $pp$ collisions in 2015. The trapping volume used here is an upgrade of the prototype which was exposed in 2012~\cite{MoEDAL2016}. It consists of 672 square aluminium rods with dimension 19$\times$2.5$\times$2.5~cm$^3$ for a total mass of 222~kg in 14 stacked boxes which were placed 1.62~m from the IP8 LHC interaction point under the beam pipe on the side opposite to the LHCb detector. 

A crucial underlying assumption for the effectiveness of the trapping technique using aluminium elements is that there is a strong binding of a magnetic monopole to the $^{27}_{13}$Al nucleus. Binding is expected between a magnetic monopole carrying the Dirac charge or higher and nuclei with non-zero magnetic moments. Existing models, summarised in Ref.~\cite{Gamberg2000}, estimate that binding should occur for $^{27}_{13}$Al (100\% natural abundance). With its large magnetic moment, $^{27}_{13}$Al has a predicted monopole-nucleus binding energy in the range $0.5-2.5$ MeV~\cite{Goebel1983,Bracci1984,Olaussen1985,Gamberg2000,Milton2006}, comparatively higher than the predictions obtained with other materials ($0.05-1$~MeV for protons, and 0.006~MeV for $^{113}_{48}$Cd). We also note that aluminium does not present a problem with respect to induced radioactivity, while its non-magnetic nature favours the stability of the SQUID magnetometer measurements. 

The samples were individually scanned with DC SQUID long-core magnetometer (2G Enterprises Model 755) newly installed at the Laboratory for Natural Magnetism at ETH Zurich. Conveniently, the new instrument features a conveyor tray for transporting samples through the sensing coils. The current induced in the superconducting coil perpendicular to the shaft is directly proportional to the magnetic flux difference in the direction of transport. A magnetic monopole contained in a sample would induce a current proportional to the pole strength. In this search, the magnetometer output is multiplied by a calibration factor to translate it into the magnetic charge contained in the sample in units of Dirac charge $g_{\rm D}$. The calibration is performed using two independent methods, as described in Ref.~\cite{DeRoeck2012b}: the solenoid method and the convolution method. For the solenoid method, small currents are flown through a physical 25~cm long solenoid of pole strength 32.4~$g_{\rm D}/\mu A$ with only one end introduced through the sensing coil. For the convolution method, a sample of known magnetic dipole moment is measured at many different positions along the magnetometer and the superposition principle is used to infer the response from a single pole. The two methods give a calibration constant that is consistent to within 10\%. The linearity of the magnetometer response is directly demonstrated for magnetic poles in the range $0.3-10^6g_{\rm D}$. 

\begin{figure*}[tb]
  \begin{center}
    \includegraphics[width=1\linewidth]{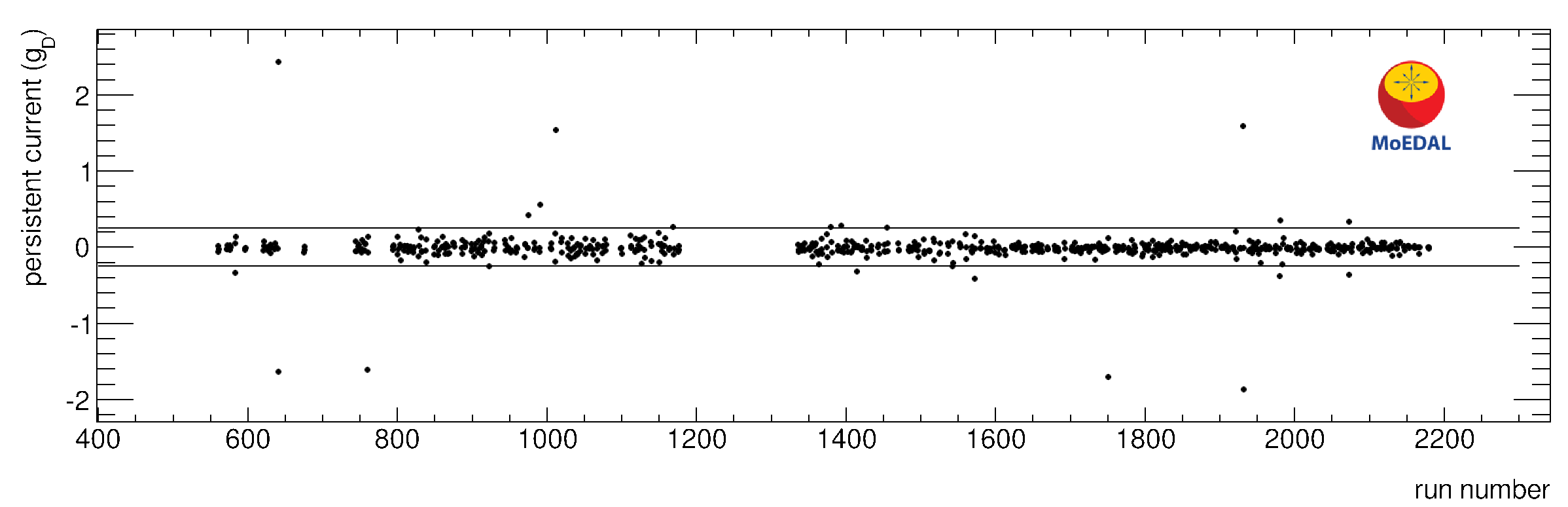}  
    \includegraphics[width=1\linewidth]{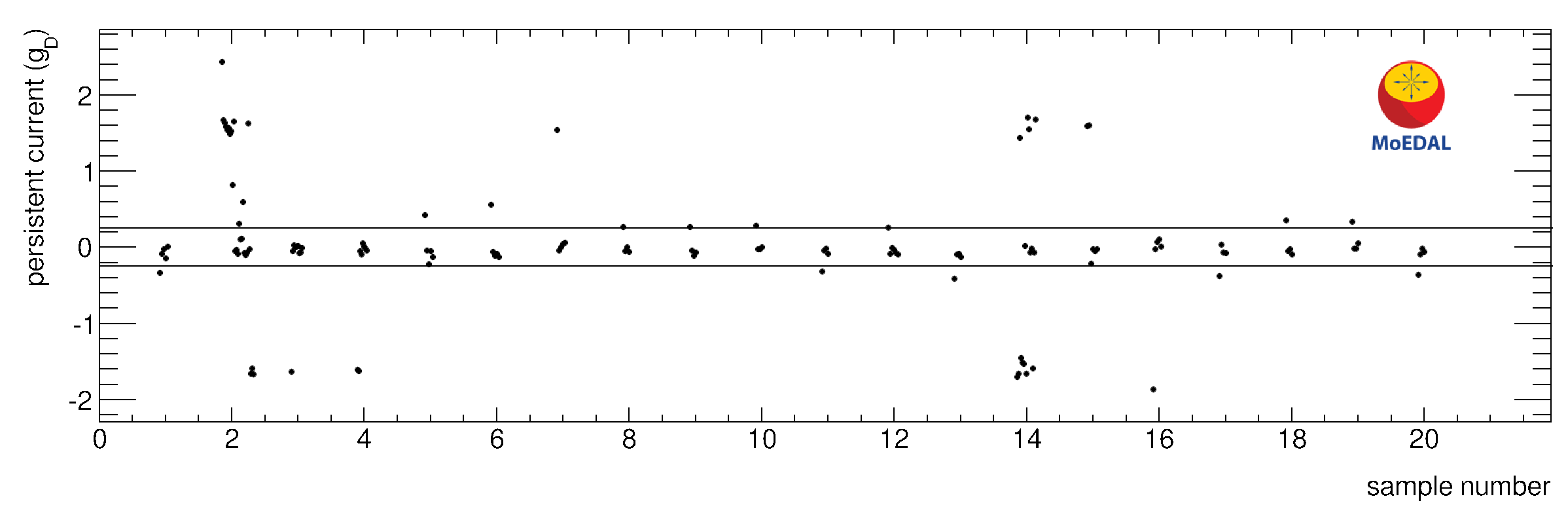}  
  \end{center}
  \caption{\label{fig:SQUID} Top: persistent current (in units of $g_{\rm D}$ after application of a calibration constant) after first passage through the magnetometer for all samples. Bottom: results of repeated measurements of candidate samples with absolute measured values in excess of $0.25g_{\rm D}$. }
\end{figure*}

The monopole signature is quantified in terms of a quantity called persistent current, defined as the difference between the induced currents measured after and before passage of the sample, to which the contribution of the conveyor tray is subtracted. The currents induced by the north and south poles of any dipole present in the sample cancel out. The persistent current is directly proportional to the magnetic charge contained in the sample and is measured in units of $g_{\rm D}$ after applying the calibration constant. Persistent currents measured during the first passage of the 672 forward trapping detector samples through the SQUID magnetometer are shown in the top panel of Fig.~\ref{fig:SQUID}. The 20 samples which yielded an absolute value corresponding to a magnetic charge larger than $0.25g_{\rm D}$ were set aside and remeasured at least 3 more times.

The bottom panel of Fig.~\ref{fig:SQUID} shows the results of the multiple measurements for these candidates. Most of these candidates are found to have a magnetisation higher than average. The magnetic dipole moments of the aluminium samples are due to impurities and correspond usually to around $5\cdot 10^{-9}$~Am$^2$ (corresponding to a response of $\sim 20g_{\rm D}$ after calibration when the sample is in the sensing region). Samples with higher magnetisation are known to cause magnetometer instabilities~\cite{DeRoeck2012b,Bendtz2013,MoEDAL2016}. Samples with a total dipole moment exceeding $1.5\cdot 10^{-7}$~Am$^2$ (corresponding to a response $> 600g_{\rm D}$ when the sample is in the sensing region) generate a flux variation corresponding to several quantum steps inside the SQUID, in which case the response may not quite come back to the same level during the flux change in the other direction as the sample passes through the sensing coil, causing a slight offset. This is the case for samples 2 and 14 for which fake signals are observed around the value $\pm 1.6g_{\rm D}$, as can be seen in Fig.~\ref{fig:SQUID} (bottom). Similar effects are found in measurements with non-exposed samples of similar or greater magnetisation (see also Refs.~\cite{DeRoeck2012b,Bendtz2013}). However, in these samples, the polarity of the persistent current depends on the sample orientation (which end of the bar is introduced first), and a fraction of the measurements still yield zero persistent current. Neither of these two features would be present in the case of a genuine magnetic monopole. Samples of weaker magnetic dipole moments consistently yield persistent currents smaller than $0.25g_{\rm D}$ in absolute value. The monopole hypothesis is thus excluded for all 20 candidates. 

There are two instances in which a genuine monopole in a sample which is measured only once could possibly yield a persistent current lower than $0.25g_{\rm D}$ such as to avoid being selected as a candidate and escape detection. The first possibility is that there exists another monopole of opposite charge in the same sample. Due to the large number of samples (672 in total), this possibility is ruled out by the following argument: in the case of a very small monopole production cross section, the probability that both the monopole and the anti-monopole (which are almost never produced in the same direction) would end up in the same sample is negligible; and in the case of a large production cross section, the probability that the positive charges would cancel the negative charges in all samples in which monopoles would be trapped is again negligible. The second possibility is that of a mismeasurement leading by chance to a cancellation of the persistent current induced by the monopole in such a way that the measured value would be in the interval $\left[-0.25g_{\rm D},0.25g_{\rm D}\right]$. This is studied using the distribution of persistent currents obtained in samples without monopoles, assuming that the magnetic field of the monopole itself (small compared to those of magnetic dipoles contained in the sample and tray) does not affect the mismeasurement probability. This probability is estimated to be less than 0.5\% for monopoles with absolute charge $0.5g_{\rm D}$ or higher. Thus, the presence of a monopole with absolute magnetic charge exceeding $0.5g_{\rm D}$ is excluded at more than 99.5\% confidence level in all samples. 

The trapping detector acceptance is defined as the probability that a monopole of given mass, charge, energy and direction would end its trajectory inside the trapping volume. It is determined from the knowledge of the material traversed by the monopole~\cite{MoEDAL2016} and the ionisation energy loss of monopoles when they go through matter~\cite{Ahlen1978,Ahlen1980,Ahlen1982,Cecchini2016} implemented in a simulation based on G\textsc{eant}4~\cite{Geant42006}. Simulations with uniform monopole energy distributions allow identification, for various charge and mass combinations, of ranges of kinetic energy and polar angle for which the acceptance is relatively uniform, called fiducial regions. The fiducial regions given in Ref.~\cite{MoEDAL2016} can conservatively be used to provide an interpretation which does not depend on the monopole production model.

\begin{figure*}[tb]
  \begin{center}
    \includegraphics[width=0.495\linewidth]{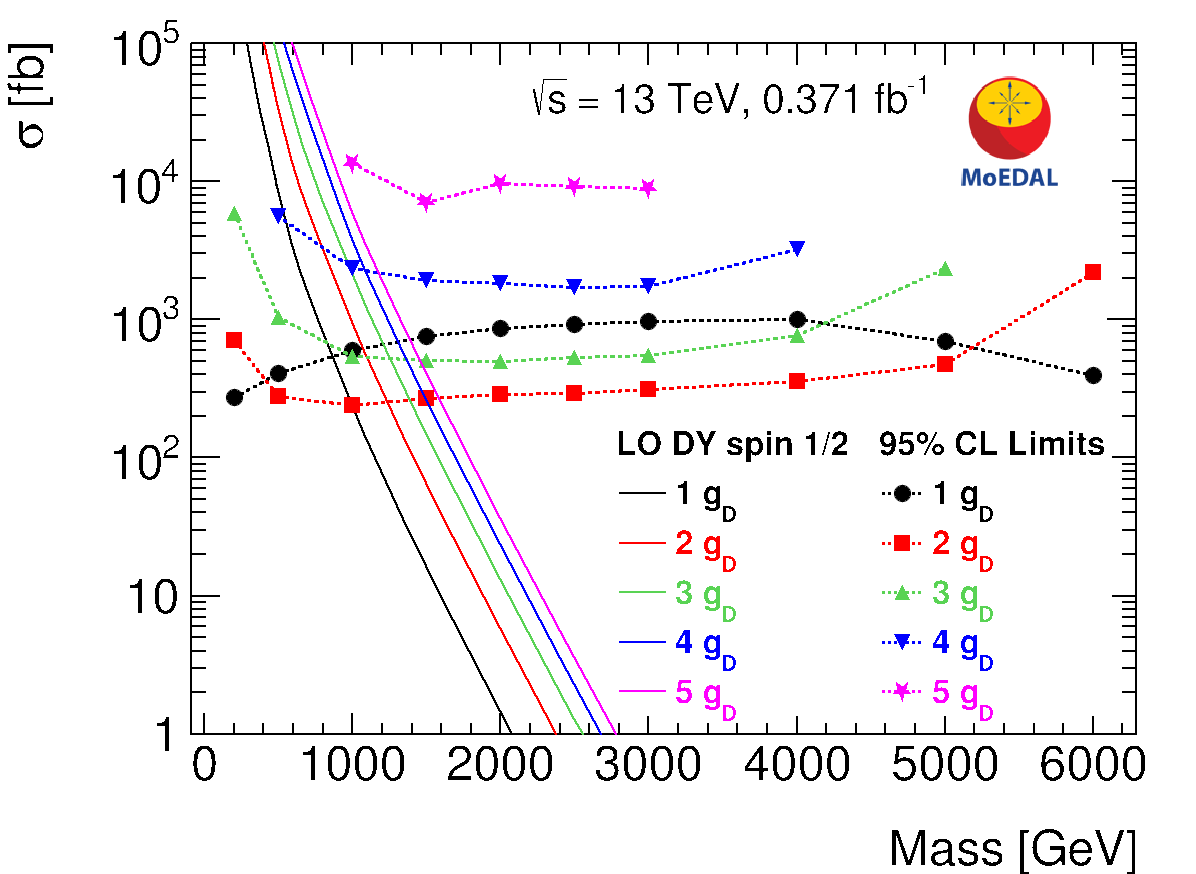}  
    \includegraphics[width=0.495\linewidth]{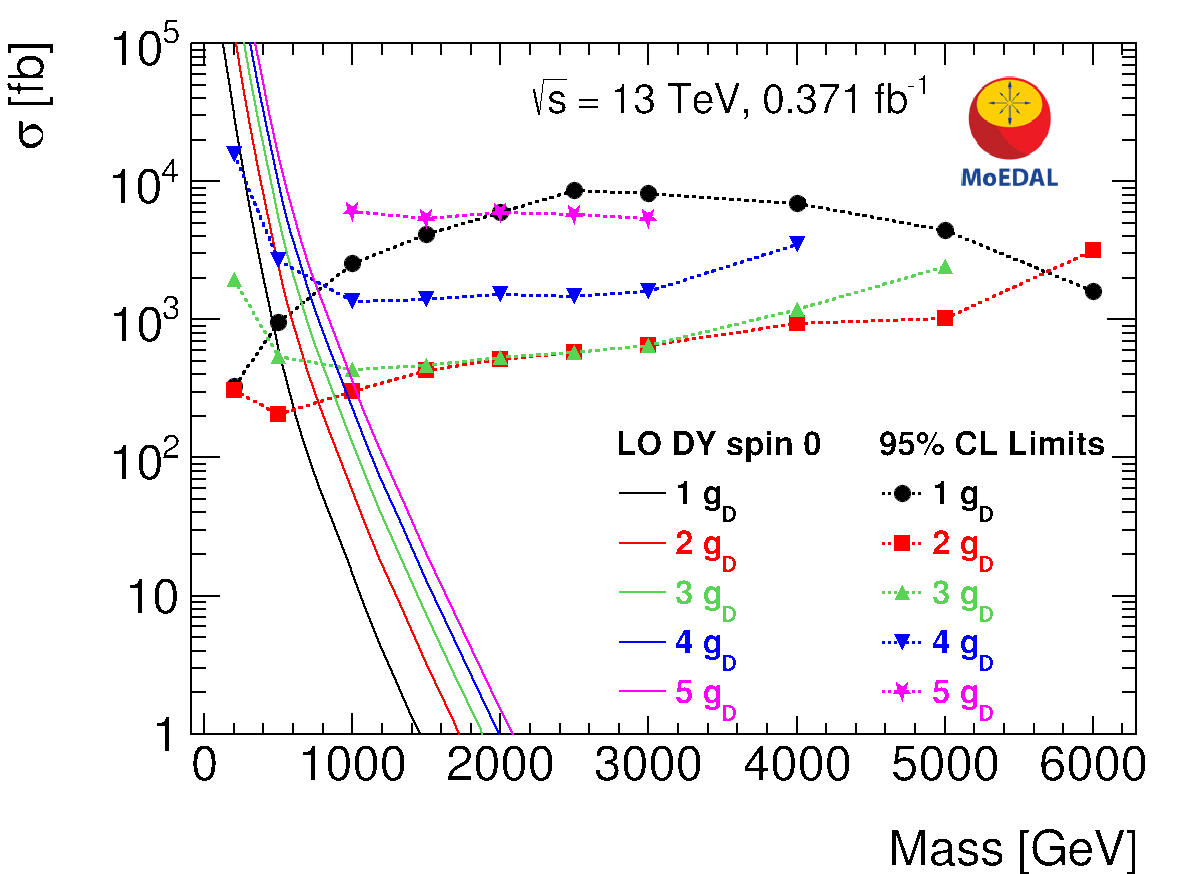}  
  \end{center}
  \caption{\label{fig:limits} Cross-section upper limits at 95\% confidence level for DY monopole production in 13 TeV $pp$ collisions as a function of mass for spin-1/2 (left) and spin-0 (right) monopoles. The colours correspond to different monopole charges. Acceptance loss is dominated by monopoles punching through the trapping volume for $|g|=g_{\rm D}$ while it is dominated by stopping in upstream material for higher charges, explaining the shape difference. The solid lines are DY cross-section calculations at leading order. }
\end{figure*}

As a realistic model of monopole pair production, a Drell-Yan (DY) mechanism is employed for its simplicity and for ease of comparison with previous LHC results~\cite{ATLAS2012a,ATLAS2015a,MoEDAL2016}. It should be noted, though, that the DY model does not constitute a reliable tool for calculating the monopole production cross section as a consequence of the fact that it is perturbative, and thus not strictly applicable to the non-perturbative regime of strong magnetic charges. Two scenarios for monopole spin are adopted, namely spin-0 and spin-1/2. Monopole trapping acceptances in the range 0.1\% to 4\% are are obtained from full G\textsc{eant}4 simulations of DY events in 13 TeV $pp$ collisions generated with {\sc MadGraph5}~\cite{Alwall2014} in the intervals $1-6g_{\rm D}$ and $200\leq M \leq 6000$~GeV, with 100000 events for each mass-charge combination. 

Acceptance loss comes from a combination of monopoles punching through the trapping volume (mostly for $|g|=g_{\rm D}$) and monopoles ranging out before reaching the trapping volume (for the higher charges). The latter effect decreases the acceptance for DY monopoles with increasing charge and reaches below 0.1\% for a charge of $6g_{\rm D}$, in which case the DY interpretation ceases to be meaningful because the systematic uncertainties exceed 100\%. The behaviour of the acceptance as a function of mass has two contributions: the mass dependence of the DY kinematic distributions (more central and less energetic monopoles at high mass), and the velocity dependence of the energy loss (lower at lower velocity for monopoles). For monopoles with $|g|=g_{\rm D}$, acceptance losses predominantly come from punching through the trapping volume, and therefore the acceptance is highest at very high masses (low initial kinetic energy from DY kinematics) and at low masses (high energy loss), with a minimum around 3000~GeV. The reverse is true for monopoles with $|g|>g_{\rm D}$ which predominantly stop in the upstream material and for which the acceptance is highest for intermediate masses. The spin dependence is solely due to the different event kinematics (more central and more energetic monopoles for spin-0). Uncertainties in the acceptance include event statistics as well as the effects of uncertainties in energy loss calculations, detector position, and material budget. The latter represents the main contribution to the acceptance uncertainty and is estimated using full simulations of monopole propagation through the setup with two additional geometries covering conservative uncertainties on the material placed upstream of the trapping detector, as described in Ref.~\cite{MoEDAL2016}. 

\begin{table}[tb]
\begin{center}
\begin{tabular}{|l|c|c|c|c|}
\hline
mass limits [GeV] & $1g_{\rm D}$ & $2g_{\rm D}$ & $3g_{\rm D}$  & $4g_{\rm D}$ \\
\hline
MoEDAL 13 TeV &&&& \\
(this result) &&&& \\
 DY spin-1/2             & 890      & 1250     & 1260  & 1100   \\
 DY spin-0               & 460      & 760      & 800   & 650    \\
\hline
MoEDAL 8 TeV &&&&  \\
 DY spin-1/2             & 700      & 920      & 840   & -- \\
 DY spin-0               & 420      & 600      & 560   & -- \\
\hline
ATLAS 8 TeV &&&&  \\
 DY spin-1/2             & 1340     & -- & -- & -- \\
 DY spin-0               & 1050     & -- & -- & -- \\
\hline
\end{tabular}
\caption{Monopole lower mass limits (95\% confidence level) in models of spin-1/2 and spin-0 DY pair production in LHC $pp$ collisions for monopole charges $|g|$ up to $4g_{\rm D}$. These limits are based upon cross sections computed at leading order. These cross sections are only indicative since the monopole coupling to the photon is too large to allow for perturbative calculations. Previous results obtained in 8 TeV collisions are from Ref.~\cite{MoEDAL2016} (MoEDAL prototype trapping detector) and Ref.~\cite{ATLAS2015a} (ATLAS).}
\label{tab:masslimits}
\end{center}
\end{table}

Cross-section limits for spin-1/2 and spin-0 monopole production are shown in Fig.~\ref{fig:limits}. They are extracted from the following inputs: the acceptance estimates and their uncertainties, assuming DY kinematics; the  luminosity of $0.371\pm 0.004$~fb$^{-1}$ obtained during the 2015 exposure to 13~TeV $pp$ collisions; the expectation of strong binding to aluminium nuclei~\cite{Milton2006} of monopoles with velocity $\beta =\frac{v}{c} \le 10^{-3}$, where $v$ is the velocity of the monopole; and the non-observation of magnetic charge inside the trapping detector samples. 

The DY cross sections computed at leading order are shown as solid lines in Fig.~\ref{fig:limits}, with the caveat, as already mentioned, that the coupling of the monopole to the photon is so large that perturbative calculations are not expected to be reliable. Spin-1/2 monopole pairs have two possible states while spin-0 pairs have only one, explaining a factor 2 in the cross section. Furthermore, the photon has spin-1, thus forcing spin-0 monopoles to carry angular momentum in the final state, reducing the cross section and leading to different kinematics. Using these cross sections and the limits set by the search, indicative mass limits are extracted and reported in Table~\ref{tab:masslimits} for magnetic charges up to $4g_{\rm D}$. For $5g_{\rm D}$ (pink stars in Fig.~\ref{fig:limits}), the trapping acceptance for masses below 1000~GeV is not good enough to set a mass limit. 

In summary, the aluminium elements of the MoEDAL trapping detector exposed to 13~TeV LHC collisions in 2015 were scanned using a SQUID-based magnetometer for the presence of trapped magnetic charge, and none were found. Consequently,  monopole-pair direct production cross-section limits in the range $200-10000$~fb were set for magnetic charges up to $5g_{\rm D}$ and masses up to 6~TeV. In a DY model with spin-1/2 monopoles, this translates into monopole mass limits exceeding 1~TeV -- the strongest to date at a collider experiment~\cite{PDG2016} -- for charges ranging from two to four times the Dirac charge. 

We thank CERN for the very successful operation of the LHC, as well as the support staff from our institutions without whom MoEDAL could not be operated efficiently. We would like to acknowledge the invaluable assistance of members of the LHCb Collaboration, in particular Guy Wilkinson, Rolf Lindner, Eric Thomas, and Gloria Corti. We thank M.~King and R.~Staszewski for their help with the software. Computing support was provided by the GridPP Collaboration~\cite{GridPP2006,Britton2009}, in particular from the Queen Mary University of London and Liverpool grid sites. This work was supported by a fellowship from the Swiss National Science Foundation; by the UK Science and Technology Facilities Council (STFC), via the research grants ST/L000326/1, ST/L00044X/1 and ST/N00101X/1; by the Spanish Ministry of Economy and Competitiveness (MINECO), via the grants Grants No. FPA2014-53631-C2-1-P and FPA2015-65652-C4-1-R; by the Generalitat Valenciana via the Projects PROMETEO-II/2013/017 and PROMETEO-II/2014/066, and by the Severo Ochoa Excellence Centre Project SEV-2014-0398; by the Physics Department of King's College London; by a Natural Science and Engineering Research Council of Canada via a project grant; by the V-P Research of the University of Alberta; by the Provost of the University of Alberta; by UEFISCDI (Romania); and by the INFN (Italy).

%

\bibliography{MMT2016}

\end{document}